\documentclass[prl,aps,showpacs,showkeys]{revtex4}

\def\XXint#1#2#3{{\setbox0=\hbox{$#1{#2#3}{\int}$}
     \vcenter{\hbox{$#2#3$}}\kern-.5\wd0}}

\usepackage{graphicx}
\usepackage{bm}
\usepackage{amssymb}
\usepackage{amsmath}
\usepackage{euscript}

\begin{document}

\title{On the Anomalous Flicker Noise Intensity in High-Temperature Superconductors}

\author{Kirill~A.~Kazakov}

\affiliation{Department of Theoretical Physics,
Physics Faculty,\\
Moscow State University, $119991$, Moscow, Russian Federation}

\begin{abstract}
The problem of anomalously high levels of flicker noise observed in the normal state of the high-temperature superconductors is addressed. It is argued that the anomaly is the result of incorrect normalization of the power spectra according to the Hooge formula. A careful analysis of the available experimental data is given, which shows that the scaling of the spectral power with sample size is essentially different from the inverse proportionality. It is demonstrated that the measured spectra obey the law given by the recently proposed quantum theory of fundamental flicker noise.
\end{abstract}

\maketitle

As is well-known, power spectra of voltage fluctuations in all conducting materials exhibit a universal low-frequency behavior called $1/f$ (flicker) noise. The origin of this noise is still a matter of controversy, the main difficulty for theoretical explanation being the observed unboundedness of the $1/f$-spectrum. There are several models of flicker noise generation, based on the usual physical mechanisms such as the defect motion \cite{kogan}, temperature fluctuations \cite{voss1}, fluctuations of the charge carrier mobility \cite{hooge1,hooge2}, or of the charge carrier number \cite{mcwhorter}. These models, however, are able to explain the inverse frequency dependence only in narrow bands covering 1 to 3 frequency decades. For instance, according to Ref.~\cite{stephany}, the noise caused by the defect motion in carbon conductors is characterized by the power spectrum close to $1/f$ only in the frequency range $10^3~{\rm Hz}$ to $10^4~{\rm Hz}.$ At the same time, the flicker noise has been detected in the range as wide as 12 decades at least -- $10^{-6}~{\rm Hz}$ to $10^{6}~{\rm Hz}.$

Another point of a long debate is the question whether this noise is a bulk effect.
It is ascertained that for a given material, intensity of $1/f$-noise produced by a sample increases as the sample dimensions decrease, but specific form of this dependence is not well-established. The famous Hooge's empirical law \cite{hooge1} states that in the case of metals, the power spectrum of voltage fluctuations, $S(f),$ is inversely proportional to the number of free charge carriers, $N,$
\begin{eqnarray}\label{hooge}
S(f) = \frac{\alpha V^2}{f N}\,,
\end{eqnarray}\noindent where $V$ is the voltage bias \footnote{Evidently, if conductance fluctuations were the noise source, then their power spectrum would be proportional to $V^2$ (in the linear regime). The fact that the measured $S(f)$ is proportional to $V^2$ is often used to reverse this statement.}, and $\alpha$ a coefficient (Hooge's constant) of the order $10^{-3}.$ Equivalently, one can say that for a fixed concentration of charge carriers, $n,$ $S(f)$ is inversely proportional to the sample volume $\Omega = N/n.$ Since the time the formula (\ref{hooge}) was proposed it has been shown experimentally that even for pure metals, $S(f)$ does not always scale as the inverse sample volume \cite{wong1}, but it is widely accepted that this dependence is ``close'' to $\Omega^{-1}.$ It is also known very well that the $1/f$ spectrum is actually $1/f^{\gamma},$ where the constant $\gamma$ is ``close'' to unity \footnote{The adjective close is taken here in quotes because its precise meaning can only be determined by an underlying theory. Indeed, it is meaningless to say that $f^{0.01} \approx 1,$ even if $f\approx 1$Hz.}. Under the assumption that deviations of the exponents from unity do not break qualitative validity of Eq.~(\ref{hooge}), the coefficient $\alpha$ in this formula has become widely used as a measure of the sample quality, the lower values implying the better noise characteristic, $10^{-3}$ being the reference value for a ``good'' material.

It came as a surprise when it was discovered \cite{testa,maeda,black2} in the late eighties that high-$T_c$ superconductors in the normal state are characterized by extremely high levels of $1/f$-noise, typically 7 to 10 orders of magnitude larger than in conventional materials. Much effort was spent to show that this effect is not an artifact of the sample preparation technique, and the noise level is practically the same in single crystals and polycrystals of the same size \cite{maeda,song1,song2,song3}. However, it was found subsequently \cite{liu,scouten} that thin-film microbridges composed of YBa$_2$Cu$_3$O$_x$ deposited on various oxide substrates exhibit much lower noise than previously reported (by at least 3 orders). The authors \cite{liu,scouten} attribute this noise reduction to the higher quality of their samples. Earlier, similar reduction was observed \cite{misra} in thin films of Tl$_2$Ba$_2$Ca$_2$Cu$_2$O$_8.$

Although some tentative arguments has been put forward to explain the anomalous noise levels in various particular cases, no physical mechanism able to account for the huge difference of the noise intensity produced by apparently similar materials has been identified. The purpose of this Letter is to explain this phenomenon on the basis of the recently developed \cite{kazakov} quantum theory of the fundamental flicker noise \footnote{The term fundamental refers to the true flicker noise, that is the one characterized by the spectrum $1/f^{\gamma}$ extending down to $f=0,$ {\it i.e.,} without a low-frequency cutoff. As was shown in \cite{kazakov}, this unboundedness of the quantum noise spectrum is consistent with finiteness of the total noise power.}. It will be shown that there is actually no anomaly in the noise levels detected in the experiments cited above, and that the found discrepancy in the orders of the $\alpha$-parameter is solely due to incorrect use of the Hooge formula for normalization of the power spectra. The main point to be proved below is that $S(f)$ does not scale as the inverse volume (or inverse charge carrier number), and its scaling law is not even ``close'' to the inverse dependence. In brief, this apparent contradiction with the general opinion can be explained as follows. In order to establish the $\Omega^{-1}$-dependence, one has to perform a series of experiments with samples of significantly different size, which are usually sufficiently small in one or two dimensions (films, bridges {\it etc.}). The problem is that varying the sample thickness is technically much more difficult than varying its length. Moreover, measurements of the dependence of $S(f)$ on the sample thickness usually give highly scattered or even non-reproducible results. By this reason, investigations of the scaling of power spectra with the sample size are mostly ``one-dimensional.'' If $S(f)$ were inversely proportional to the sample volume, then it would be also inversely proportional to its length. Although the converse is not true, it is often used to assert the $\Omega^{-1}$-dependence.

According to \cite{kazakov}, the fundamental flicker noise power spectrum has the form
\begin{eqnarray}\label{mainhom}
S(f) = \frac{\varkappa V^2}{f^{1 +
\delta}}\,, \quad \varkappa \equiv \frac{e^3c^{\delta}}{\pi\hbar^2
c^3}\mu T g\,.
\end{eqnarray}
\noindent In this formula, $\mu$ is the charge carrier mobility in the direction of the external electric field, $T$ the absolute temperature, $e$ the elementary charge, $c$ the speed of light, and $g$ a geometrical factor. In the practically important case when the sample is an elongated parallelepiped, one has for $\delta > 0,$
\begin{eqnarray}\label{gfactorapprox1}
g = \frac{2}{\delta \EuScript{D}(\EuScript{D} -
2)a^{2\delta}L^{1-\delta}}\,, \quad \EuScript{D} = 3 - \delta,
\end{eqnarray}
\noindent where $a$ is the sample thickness, and $L\gg a$ its length. In the case $\delta = 0,$ the geometrical factor takes the form $g = 2/(3L)\ln(L/w)$ ($w\ll L$ is the sample width; this expression is valid with logarithmic accuracy). In the $CGS$ system of units, the $\varkappa$-factor reads
\begin{eqnarray}\label{kappaapprox}
\varkappa \approx 1.62 \cdot 10^{10.48\,\delta - 22}g\mu T\,,
\end{eqnarray}
\noindent where $T$ is to be expressed in $^{\circ}{\rm K}.$

First of all, it is to be noted that the noise level is very sensitive to the value of $\delta = \gamma -1$: Collecting $\delta$'s in the exponents in Eqs.~(\ref{mainhom}), (\ref{gfactorapprox1}) shows that $S(f)$ is proportional to  $\left(c L/fa^2\right)^{\delta}\,.$ It is a common situation in the experimental literature that $\gamma$ is reported equal to unity, while inspection of the power spectra (if any) gives, say, $\gamma = 1.1.$ Substituting $a = 10^{-5}$cm, $L = 1$mm, $f=1$Hz in the above expression shows that the error of $0.1$ in the value of $\gamma$ gives rise to an extra factor $10^2$ in the noise magnitude. Also, it should be mentioned that dependence of $\delta$ on the sample thickness is more pronounced than its dependence on the sample length (because $\delta$ reflects the properties of the photon propagator in the sample, which are sensitive to the boundary conditions -- the type of substrate, surface roughness {\it etc.}). Presumably, this is why experiments show significant scatter in the dependence of $S(f)$ on $a.$ Parameters entering Eq.~(\ref{mainhom}) can be most accurately determined in pure metals. For instance, using the results \cite{wong1} of flicker noise measurements in copper films it was shown \cite{kazakov} that this equation agrees with the experimental data within the accuracy better than one order (the experimental error in the frequency exponent in \cite{wong1} is only $0.02$).

In the experimental data analyzed below, $\gamma$ is determined with the accuracy about $0.1,$ which as we just saw implies a two-order ambiguity in the noise magnitude in thin films, or somewhat smaller in thicker samples. Fortunately, this is sufficient for our analysis because the discrepancy in the values of the Hooge constant is much larger -- up to 10 orders. Unfortunately, not all of the papers cited above provide information needed in estimating the noise level according to Eq.~(\ref{mainhom}). The most important parameters are the sample thickness and length, and the frequency exponent. The sample thickness is not specified in Refs.~\cite{testa,maeda}, sample length -- in Ref.~\cite{black2}, while  Ref.~\cite{song3} gives neither sample size nor the frequency exponent. In what follows, the charge carrier mobility is estimated using the relation $\mu = 1/(\rho e n),$ where $\rho$ is the sample resistivity, and $n$ the concentration of the charge carriers, taken equal to $10^{21}$cm$^{-3}$ in all cases.

A comprehensive investigation of $1/f$-noise in bulk materials composed of Tl$_2$Ba$_2$Ca$_{k-1}$Cu$_k$O$_{4+2k}$ was carried out in Ref.~\cite{song1}. This paper reports the results of noise measurements on four representative samples with $k=2,3$ and different preparation histories. All samples had $a = 0.1$ cm, $w = 0.3$ cm, and the distance between the voltage leads $L = 0.6$ cm. The sample resistivities are specified at $T = 200$ K in Table~III of \cite{song1}. As to the noise exponent, the authors mention that for all samples, $\gamma = 1.08 \pm 0.1$ for all normal state temperatures. However, an example of the power spectrum given for one of the samples (sample 4) on Fig.~2 clearly shows that in this case, $\gamma$ does not belong to the declared interval, being equal approximately to $1.29$ at $T = 295$ K. To give a theoretical estimate of the noise level, we take the mean value $\gamma = 1.08$ for the first three samples, $\gamma = 1.29$ for the fourth, and use the above data to evaluate $\mu,g,\varkappa.$ The results of the calculation, together with the experimental values $\varkappa_{\rm exp}$ inferred from Figs.~3,4 of Ref.~\cite{song1}, are summarized in Table~\ref{tsong1}. Barring errors in determining the charge carrier mobility,  inaccuracy of the calculated $\varkappa,$ corresponding to the error $0.1$ in $\delta,$  is a factor $\left(3\cdot 10^{10}0.6/0.1^2\right)^{0.1} \approx 17.$ It is seen from the table that the calculated and measured values of $\varkappa$ agree well within this accuracy.

Soon after the work \cite{song1}, considerably lower values of $\alpha$ (close to those in conventional metals) were detected \cite{misra} in thin films of Tl$_2$Ba$_2$Ca$_2$Cu$_2$O$_8.$ The films had $a = 1$ $\mu$m, and $\gamma$ was loosely indicated to be in the range $1.10 \pm 0.15$ for all temperatures. The other sample dimensions are also loosely described to be typically $L = 300$ $\mu$m, $w = 25$ $\mu$m. The resistivity data is given only for sample A (see Fig.~1 of Ref.~\cite{misra}). We use this data to calculate $\varkappa$ for four different temperatures in the normal state. $\delta$ is taken equal to $0.1$ in all cases. The results of the calculation are compared in Table~\ref{tmisra} with the experimental values $\varkappa_{\rm exp}$ inferred from Fig.~2 of Ref.~\cite{misra}. Because of the small sample thickness and the relatively large indeterminacy in $\gamma,$ the accuracy of $\varkappa_{\rm th}$ is a factor $\left(3\cdot 10^{10}0.03/(10^{-4})^2\right)^{0.15} \approx 350.$ As is seen from Table~\ref{tmisra}, however, the calculated and measured values of $\varkappa$ actually agree up to one order of magnitude.

$1/f$-noise measurements in bulk samples of YBa$_2$Cu$_3$O$_x$ are described in Ref.~\cite{song2}. The samples used were single crystals with $L = w = 0.2$ cm, $a = 0.01$ cm. Two types (A and B) of crystals were considered, with and without superconducting state, depending on the degree of oxygenation. The frequency exponent is again loosely determined to belong to the interval $1.06 \pm 0.1.$ Using the resistivity and power spectrum data from Figs.~1, 2 of Ref.~\cite{song2}, one obtains theoretical estimates of the $\varkappa$-parameter and other relevant parameters as shown in Table~\ref{tsong2}. In the present case, inaccuracy of the calculated $\varkappa,$ corresponding to the error $0.1$ in $\delta,$ is a factor $\left(3\cdot 10^{10}0.2/0.01^2\right)^{0.1} \approx 25.$ In particular, the seemingly large discrepancy in the calculated and measured values of $\varkappa$ in the case of the sample A corresponds to the error of only $0.15$ in the frequency exponent, so that we can state satisfactory agreement in this case too.

In 1994, a systematic investigation of flicker noise in thin films of YBa$_2$Cu$_3$O$_x$ deposited on LaAlO$_3$ substrate was undertaken \cite{liu} in order to determine its dependence on the oxygen content, $x.$ Unexpectedly, it was found that the values of $\alpha$ in this case, although still huge compared to pure metals, are several orders of magnitude smaller than those reported previously. All samples had $a = 8.5~10^{-6}$ cm, $L = 0.5$ cm, and $w = 0.07$ cm. The experimental data provided allows theoretical assessment of the noise intensity for the following values of the parameter $x$: $6.81, 6.62,$ and $6.43.$ The authors give for the frequency exponent in these three cases the values $\gamma = 1.0,$ $1.1$ and $1.1,$ respectively. However, careful evaluation of the slopes of the best fits drawn on Fig.~2 of \cite{liu} shows that the corresponding values are actually $1.03,$ $1.09$ and $1.14.$ Below, we use these more accurate figures to calculate $\varkappa_{\rm th}.$ The room temperature resistivities are taken from Fig.~1, while the values of $\varkappa_{\rm exp}$ are inferred from Fig.~3 of Ref.~\cite{liu}. The results of the calculation are summarized in Table~\ref{tliu}. If we assume that the experimental error in $\delta$ is about $0.04$ (which is the difference between the values for $\gamma$ given in the paper \cite{liu} and the more accurate values indicated above), then the error in the calculated $\varkappa$ is a factor $\left(3\cdot 10^{10}\cdot0.5/(8.5\cdot 10^{-6})^2\right)^{0.04} \approx 6.5.$ We see that $\varkappa_{\rm th}$ and $\varkappa_{\rm exp}$ are in a good agreement within this accuracy. One should remember that the actual error in the calculated $\varkappa$ can be somewhat higher because of the errors in other factors in Eq.~(\ref{mainhom}), primarily $\mu.$

In a later work \cite{scouten}, flicker noise measurements on thin-film microbridges of YBa$_2$Cu$_3$O$_x$ deposited on MgO and SrTiO$_3$ substrates were performed, and the effects of electromigration and oxygen plasma annealing on the noise characteristics investigated. In particular, further reduction in the values of $\alpha,$ by about three orders in comparison with Ref.~\cite{liu}, was reported.
Unfortunately, the authors of \cite{scouten} did not fully specify the size of the samples used, gave no resistivity data, and what is worse, did not mention the values of $\gamma$ at all. Yet, since the found reduction in the noise magnitude is very interesting for the present discussion, some guesswork will be done to fill up the missing information. The sample volume was specified in Ref.~\cite{scouten} to be $3\cdot 10^{-12}$cm$^3,$ and it was also mentioned that samples with $w=1, 2, 5$ $\mu$m and the length/width ratio 2.5 to 1 were used. It easy to check that the choice $w = 2$ $\mu$m, $L = 4$ $\mu$m minimizes potential errors in the sample length and thickness $a = \Omega/Lw \approx 4\cdot 10^{-5}$ cm to less than one order. As to the frequency exponent and the charge carrier mobility, we will take their most common values for YBCO films: $\delta = 0.1,$ $\mu = 1$ cm$^2$/Vs.
Calculation gives $g = 3.4\cdot 10^{4}$ units CGS, and then $\varkappa_{\rm th} = 5\cdot 10^{-12}$ for $T = 300$ K, while Fig.~1 of Ref.~\cite{scouten} shows that $\varkappa_{\rm exp} \approx 10^{-12}$ (according to \cite{scouten}, the choice of the substrate as well as the effects of electromigration and annealing do not change the order of $\varkappa_{\rm exp}$). Assuming that the error in $\delta$ is $0.1,$ the accuracy of this comparison is about two orders.

The above analysis allows us to draw the following conclusions:

1) In all cases where experimental data permits theoretical evaluation, the measured values of $\varkappa$ agree within experimental error with those calculated according to Eqs.~(\ref{mainhom}) -- (\ref{kappaapprox}). The accuracy ranges from about one order of magnitude in thick long samples to about two orders in thin films.

2) The anomalously high noise levels found in bulk samples of the copper-oxide superconductors are the result of normalization of the power spectra according to Eq.~(\ref{hooge}).

These facts naturally explain the many-order anomaly in the values of $\alpha$ found in bulk superconductors, as well as its reduction observed in thin films. The point is that the Hooge formula was initially gauged on metal films, so there is no surprise that it gives sensible (up to 1-2 orders of magnitude) predictions when applied to films made of other materials. This is because the scaling of the noise level with sample thickness is not important in this case, while the scaling with the sample length in Eq.~(\ref{hooge}) is very close to that in Eq.~(\ref{mainhom}). The difference between the two scalings with $a$ becomes pronounced in thick samples, thus leading to the observed ``anomaly'' in the noise level.

The above results also explain why $1/f$-noise in pure metals is noticeable only in sufficiently thin samples. Experiments show that the frequency exponent in conventional metals is normally very close to 1 in samples with $a \gtrsim 10^{-4}$ cm, while in other materials it is often as large as $1.2-1.3$ even in $1$ mm-thick samples. At the same time, we have seen that $\varkappa$ in Eq.~(\ref{mainhom}) is very sensitive to the value of $\gamma,$ and the increase by $0.3$ in $\gamma$ may well give rise to several orders of magnitude in the noise level. In this connection, it should be recalled that the factors $\mu,T$ in Eq.~(\ref{mainhom}) are also important in comparing the noise levels in different materials (in semiconductors, for instance, the charge carrier mobility brings in another 2-3 orders in the noise magnitude when compared to metals).

Finally, we mention that another interesting experimental evidence can be explained at least qualitatively within the theory developed in \cite{kazakov}, namely, the noise amplification in the superconducting transition region. According to this theory, the fundamental flicker noise originates from quantum interaction of the charge carriers with the photon heat bath. In the normal state, the carriers are not correlated, and their contributions to the voltage fluctuation add up with different phases, so that the total noise intensity remains at the level of individual contribution. Things change, however, in the transition region because
of the growth of the correlation radius: various contributions add up coherently, and the total noise magnitude grows rapidly as $T \to T_c,$ before it drops down to zero at $T=T_c$ (where $V$ vanishes).

\pagebreak

\begin{table}
\begin{tabular}{cccccc}
\hline\hline
{\rm sample}
  & \hspace{0,1cm} $\delta$ \hspace{0,1cm}
  & \hspace{0,1cm} $\mu\times 10^{-2}$ \hspace{0,1cm}
  & $g$
  & \hspace{0,1cm} $\varkappa_{\rm th}\times 10^{15}$ \hspace{0,1cm}
  & $\varkappa_{\rm exp} \times 10^{15}$   \\
\hline
1& 0.08 & 1.1 & 21 & 0.5 & 1.8 \\
2& 0.08 & 13 & 21 & 6 & 25 \\
3& 0.08 & 26 & 21 & 12 & 30 \\
4& 0.29 & 5.4 & 19 &  180  & 580\\
\hline\hline
\end{tabular}
\caption{Calculated ($\varkappa_{\rm th}$) and measured
($\varkappa_{\rm exp}$) values of $\varkappa$ for the four bulk samples of Tl$_2$Ba$_2$Ca$_{n-1}$Cu$_n$O$_{4+2n}$ from Ref.~\cite{song1}. $T = 200$ K, $\mu, g ,\varkappa$ are given in the CGS system of units.} \label{tsong1}
\end{table}

\begin{table}
\begin{tabular}{cccccc}
\hline\hline
  $T,$ K
  & \hspace{0,1cm} $\delta$ \hspace{0,1cm}
  & \hspace{0,1cm} $\mu\times 10^{-3}$ \hspace{0,1cm}
  & $g\times 10^{-3}$
  & \hspace{0,1cm} $\varkappa_{\rm th}\times 10^{13}$ \hspace{0,1cm}
  & $\varkappa_{\rm exp} \times 10^{13}$   \\
\hline
150& 0.1 & 12.6 & 1.1 & 38 & 1.4 \\
200& 0.1 & 9.5 & 1.1 & 38 & 1.4 \\
250& 0.1 & 7.5 & 1.1 & 37 & 1.7 \\
300& 0.1 & 6.3 & 1.1 & 38 & 3.5 \\
\hline\hline
\end{tabular}
\caption{Same for sample A of Tl$_2$Ba$_2$Ca$_2$Cu$_2$O$_8$ thin film from Ref.~\cite{misra}.} \label{tmisra}
\end{table}

\begin{table}
\begin{tabular}{cccccc}
\hline\hline
  {\rm sample}
  & \hspace{0,1cm} $\delta$ \hspace{0,1cm}
  & \hspace{0,1cm} $\mu\times 10^{-2}$ \hspace{0,1cm}
  & $g$
  & \hspace{0,1cm} $\varkappa_{\rm th}\times 10^{14}$ \hspace{0,1cm}
  & $\varkappa_{\rm exp} \times 10^{14}$   \\
\hline
A& 0.06 & 4.3 & 94 & 1 & 140 \\
B& 0.06 & 13.5 & 94 & 2.6 & 14 \\
\hline\hline
\end{tabular}
\caption{Same for the bulk samples of YBa$_2$Cu$_3$O$_x$ from Ref.~\cite{song2}. $T = 300$ K.} \label{tsong2}
\end{table}

\begin{table}
\begin{tabular}{cccccc}
\hline\hline
  \hspace{0,2cm}$x$ \hspace{0,2cm}
  & \hspace{0,3cm} $\delta$ \hspace{0,3cm}
  & \hspace{0,1cm} $\mu\times 10^{-2}$ \hspace{0,1cm}
  & $g$
  & \hspace{0,1cm} $\varkappa_{\rm th}\times 10^{14}$ \hspace{0,1cm}
  & $\varkappa_{\rm exp} \times 10^{14}$   \\
\hline
6.81& 0.03 & 62 & 46 & 3 & 2 \\
6.62& 0.09 & 19 & 65 & 5 & 1 \\
6.43& 0.14 & 6.2 & 140 & 12 & 3 \\
\hline\hline
\end{tabular}
\caption{Same for the YBa$_2$Cu$_3$O$_x$ films from Ref.~\cite{liu}. $T = 300$ K.} \label{tliu}
\end{table}


\begin{thebibliography}{}


\bibitem{kogan}
Sh.M.~Kogan and K.F.~Nagev, Sov. Phys. Solid State {\bf 24}, 1921 (1982);
R.D.~Black, P.J.~Restle, and M.B.~Weissman, Phys.~Rev.~Lett. {\bf 51}, 1476 (1983); S.~Feng, P.A.~Lee, and A.D.~Stone, Phys.~Rev.~Lett. {\bf 56}, 1960 (1986).

\bibitem{voss1}
R.F.~Voss and J.~Clarke, Phys.~Rev.~B{\bf 13}, 556 (1976);
J.~Clarke and T.Y.~Hsiang, Phys.~Rev.~Lett. {\bf 34}, 1217
(1975).

\bibitem{hooge1}
F.N.~Hooge, Phys.~Lett A {\bf 29}, 139 (1969).

\bibitem{hooge2}
F.N.~Hooge, Physica (Utr.) {\bf 60}, 130 (1972); F.N.~Hooge and A.M.H.~Hoppenbrouwers, Physica (Utr.) {\bf 45 }, 386 (1969); Th.G.M.~Kleinpenning and D.~A.~Bell, Physica (Utr.) {\bf 81B},
301 (1976); L.K.J.~Vandamme and Gy.~Trefan, Fluctuation and Noise Lett. {\bf
1}, R175 (2001).

\bibitem{mcwhorter}
A.L.~McWhorter,  In Semiconductor Surface Physics, ed.
R.H.~Kingston (University of Pennsylvania, Philadelphia, 1957),
p. 207; A.~van der Ziel, Appl.~Phys.~Lett. {\bf 33}, 883 (1978);

\bibitem{stephany}
J.F.~Stephany, J.~Phys.:~Condens.~Matter {\bf 12}, 2469 (2000).

\bibitem{wong1}
H.~Wong, Y.C.~Cheng, and G.~Ruan, J.~Appl.~Phys. {\bf 67}, 312
(1990); H.~Wong, Microelectron.~Reliab. {\bf 43}, 585 (2003).

\bibitem{testa}
J.A.~Testa {\it et al.}, Phys.~Rev.~B {\bf 38}, 2922 (1988).

\bibitem{maeda}
A.~Maeda {\it et al.}, Physica C {\bf 160}, 443 (1989).

\bibitem{black2}
R.D.~Black {\it et al.}, Appl.~Phys.~Lett. {\bf 55}, 2233 (1989).

\bibitem{song1}
Y.~Song {\it et al.}, Physica C {\bf 172}, 1 (1990).

\bibitem{song2}
Y.~Song {\it et al.}, Phys.~Rev.~Lett. {\bf 66}, 825 (1991).

\bibitem{song3}
Y.~Song {\it et al.}, Phys.~Rev.~B {\bf 45}, 7574 (1992).

\bibitem{liu}
Li Liu {\it et al.}, Phys.~Rev.~B {\bf 49}, 3679 (1994).

\bibitem{scouten}
S.~Scouten, Y.~Xu, B.H.~Moeckly, and R.A.~Buhrman, Phys.~Rev.~B {\bf 50}, 16121 (1994).

\bibitem{misra}
A.~Misra {\it et al.}, Appl.~Phys.~Lett. {\bf 59}, 863 (1991).

\bibitem{kazakov}
K.A.~Kazakov, Phys.~Lett.~A {\bf 372}, 749 (2008);
Physica B {\bf 403}, 2255 (2008).

\end{thebibliography}
\end{document}